\documentclass[12 pt, a4paper]{article}
\usepackage{physics}
\usepackage{jheppub}
\usepackage{bm}
\usepackage{xcolor}
\usepackage{amsmath}
\usepackage{amsfonts}
\usepackage{url}
\usepackage{graphicx}
\usepackage{ragged2e}
\usepackage{caption}
\usepackage{subcaption}
\usepackage{chngcntr}
\counterwithout{equation}{section}
\usepackage[normalem]{ulem}

\usepackage{lmodern}
\usepackage[utf8]{inputenc}
\usepackage[T1]{fontenc}
\definecolor{refkey}{gray}{0.45}
\definecolor{labelkey}{RGB}{155,48,48}
\hypersetup{
  colorlinks=true,
  citecolor=magenta,
  linkcolor=blue,
  urlcolor=violet
 }

\newcommand{\e}{\epsilon}

\newcommand{\bee}{\begin{equation}}
\newcommand{\ee}{\end{equation}}
\newcommand{\beq}{\begin{equation*}}
\newcommand{\eeq}{\end{equation*}}
\newcommand{\baa}{\begin{equation}\begin{aligned}}
\newcommand{\ea}{\end{aligned}\end{equation}}

\title{\center{ \fontfamily{lmr}\selectfont Finite Cut-Off Holography and the DBI Counter-Term}}

\author{\center \fontfamily{lmr}\selectfont Dileep P.  Jatkar\textsuperscript{\dag \ddag} and Upamanyu Moitra\textsuperscript{*}}

\affiliation{\begin{center}

\textsuperscript{\dag}Harish-Chandra Research Institute,  Chhatnag Road, Jhunsi, Allahabad 211019, India

\textsuperscript{\ddag}Homi Bhabha National Institute, Training School Complex, \\Anushaktinagar,  Mumbai 400094, India

\textsuperscript{*}Institute for Theoretical Physics,  Institute of Physics, Universiteit van Amsterdam, Science Park 904, 1089 XH
Amsterdam, The Netherlands

\href{mailto:dileep@hri.res.in}{\textup{\texttt{dileep@hri.res.in}}}\textup{,}
\href{mailto:u.moitra@uva.nl}{\textup{\texttt{u.moitra@uva.nl}}}
\end{center}}

\abstract{{\centering We demonstrate some very special features of the Dirac-Born-Infeld--like (DBI) gravitational counter-term in AdS$_4$ spacetime, in the context of holography with a sharp radial cut-off. We show that the three-sphere partition function is not only independent of a constant radial cut-off, but also remains unchanged under deformations of the cut-off surface. We also consider the renormalized holographic entanglement entropy for an equatorial Ryu-Takayanagi surface with a cut-off with an arbitrary shape and show that it can also be independent of the cut-off under a special condition. We also numerically study the behavior of the renormalized entropy with different counter-terms and relate the results to monotonicity properties under holographic renormalization group flow. The DBI counter-term is always seen to be associated with integrating out fewer degrees of freedom compared to other counter-terms.
}

}

\makeatletter
\gdef\@fpheader{}
\makeatother

\begin{document}
\maketitle

\section{Introduction}\label{sec-intro}

Entanglement entropy,  defined as the von Neumann entropy corresponding to a reduced density matrix for  a subsystem,  has been one of the key measures used for studying quantum entanglement between the subsystems of a given system.  There are other measures used for quantifying the entanglement between the parts of a system -- these include R\'enyi entropy,  relative entropy and mutual information.
The issue of entanglement takes primacy in the context of the black hole information paradox.  Hawking's observation \cite{Hawking:1976ra} that the radiation from a black hole is purely thermal indicated a non-conservation of information in quantum gravity --- this result potentially put the program of combining quantum mechanics with gravity in jeopardy.

In 1983,  Sorkin \cite{Sorkin:1984kjy} argued that the black hole horizon is a natural entangling surface for the degrees of freedom across the horizon, thereby proposing an interpretation of the Bekenstein-Hawking entropy \cite{Bekenstein:1973ur,  Hawking:1975vcx} of black holes in terms of quantum entanglement.  This was one of the early clues to reconciling quantum mechanics and general relativity.  About a decade later,  Srednicki \cite{Srednicki:1993im} provided a more general characterization of this relation between these two entropies by studying the entanglement of quantum fields across an entangling surface --- the entanglement entropy,  like black hole entropy,  was shown to follow an area law.

The holographic principle \cite{tHooft:1993dmi,Susskind:1994vu} has profoundly reshaped our understanding of quantum field theories and gravity, especially through the AdS/CFT correspondence \cite{Maldacena:1997re,Gubser:1998bc,Witten:1998qj}, which relates gravitational dynamics in asymptotically Anti-de Sitter (AdS) spacetimes to conformal field theories (CFTs) on their boundary.  This correspondence has endured several tests and has shed light on non-perturbative aspects of quantum gravity as well as quantum field theories.
One of the most important consequences of this correspondence is the Ryu-Takayanagi (RT) proposal  \cite{Ryu:2006bv,  Ryu:2006ef}, which geometrizes the entanglement entropy on the boundary theory in terms of the area of the minimal surface anchored to the AdS boundary.

However,  in the holographic computation of important field theory quantities such as partition functions,  one often encounters ultra-violet (UV) divergences, necessitating careful regularization and renormalization \cite{Skenderis:2002wp} to extract finite and physically meaningful quantities.
The regularization of the divergences in holography is achieved by considering an explicit radial cut-off, introducing a suitable covariant counter-term,  such as the Balasubramanian-Kraus (BK) counter-term \cite{Balasubramanian:1999re}  and sending the cut-off to spatial infinity, whereby one obtains finite, renormalized answers.
Different choices of boundary counter-terms correspond to different renormalization schemes and a special choice can possibly make the computation amenable to some appropriate physical interpretation.  A particularly elegant method that achieves cut-off independence in AdS$_4$ involves the introduction of the Dirac-Born-Infeld--type (DBI) boundary counter-term \cite{Jatkar:2011ue,Jatkar:2022zdz}.  Originally developed as an alternative to make the holographic stress tensor finite,  this counter-term has emerged as a versatile tool beyond its initial context.  We recently revisited the DBI counter-term with the motivation of checking its efficacy in various black hole backgrounds \cite{Jatkar:2022zdz}.  Note that the BK and DBI counter-terms agree when the cut-off is taken to infinity; however, in many contexts one needs to keep the cut-off finite and, in that case, contributions of these two counter-terms differ.  This can be easily seen by expanding the DBI counter-term for small curvatures \cite{Jatkar:2011ue}.

Just before the discovery of the AdS/CFT correspondence,  it was proposed that to describe the gauge theory dynamics one needs a geometry with one extra dimension, which plays the role of the energy scale \cite{Polyakov:1997tj}.    In the wake of the formulation of the AdS/CFT correspondence,  a relation between the UV scale in the field theory to the infra-red (IR) scale in the bulk was spelt out \cite{Susskind:1998dq} and there was a flurry of activity to establish that the extra dimension of the AdS space plays the role of the renormalization group  (RG) scale.  In \cite{deBoer:1999tgo}, the Einstein equations were recast into a form of the flow equation using the Hamilton-Jacobi method.  It was then shown to be related to the Callan-Symanzik equation for the boundary theory.  At a later stage,  a more refined description, which included multi-trace operators as well, was used to give a concrete realization of the holographic RG flow and its connection with Wilsonian RG \cite{Heemskerk:2010hk,Faulkner:2010jy, Balasubramanian:2012hb}.  In this paradigm,  one can consider a notion of holography at a finite radial cut-off (translating into a sharp UV cut-off in the boundary theory) and carry out various tests of the correspondence.  The boundary counter-terms play a crucial role in determining the RG flow.  It is worth mentioning that the BK counter-term has been related \cite{Taylor:2018xcy,  Hartman:2018tkw} to a certain field theory deformation involving the stress tensor operator in the context of finite cut-off holography; see also \cite{McGough:2016lol}.

In this note,   we will study some examples where we demonstrate the special nature of the DBI counter-term compared to any other counter-term.  In three-dimensional field theories,  the $F$-function, related to the renormalized free energy on a three-sphere,  has the conjectured property that it monotonically decreases along the RG trajectory \cite{Jafferis:2011zi}.   This monotonicity property is sometimes known as the $F$-theorem,  in analogy with the $c$- and $a$-theorems \cite{Zamolodchikov:1986gt,  Komargodski:2011vj}.  While holographic monotonicity theorems have been discussed extensively in the literature --- see \cite{Nishioka:2018khk} for a review --- we pursue a different approach. In  \S\ref{sec-s3pf}, we  first consider a candidate $F$-function,  the holographic partition function of the three-dimensional boundary theory on a sphere $S^{3}$ at a finite radial cut-off.   We compare the contribution to the $F$-function due to the BK counter-term at a finite radial cut-off with that of the DBI counter-term at the same  cut-off.  We show that the $F$-function explicitly depends on the radial cut-off in case of the BK counter-term and it monotonically decreases as one takes the cut-off deep into the AdS space.  On the contrary, the $F$-function with the DBI counter-term is independent of the radial cut-off and hence does not run as the radial cut-off is varied.  {In this paper,  we find compelling evidence for an even stronger result that the $F$-function with DBI counter-term is a topological invariant by considering a smooth, azimuthally symmetric,  but otherwise arbitrary cut-off surface with the topology of a 3-sphere. }  We show that the change in the on-shell Lagrangian is a total derivative without any boundary contribution and hence,  the $F$-function is unaffected and has a constant value for surfaces of aforementioned nature.  We expect that this result would extend to any arbitrary smooth surface with the same topology.

We then consider in \S\ref{sec-rtdbi} the computation of holographic entanglement entropy in presence of the DBI counter-term.   The area of the RT surface diverges when the cut-off surface is taken to infinity,  reproducing the expected area-law divergence in field theory.   One can define a notion of renormalized entanglement entropy (REE) \cite{Taylor:2016aoi} for arbitrary boundary counter-terms by using the replica method on the gravitational path integral \cite{Lewkowycz:2013nqa}.
In the case of a planar boundary,  the BK and DBI counter-terms have identical contributions essentially due to the flatness of the boundary manifold.  We then consider the entangling surface to be the equatorial circle on the boundary of the spatial section of global AdS$_{4}$ --- in this case we find that the REE with the BK counter-term is dependent on the cut-off,  whereas it is independent of the cut-off in the DBI case.  We also show that this result continues to hold if the equatorial RT surface lands orthogonally on any arbitrary cut-off surface.

We then take up the case of the entangling circle on the boundary sphere being smaller than the equator.  The smaller of the two regions separated by the entangling surface is sometimes referred to as the ``polar cap'' region in the literature.  In this case,  we need to take recourse to numerical computations and we find that when we change the radial cut-off or we change the size of the circle,  in both cases the DBI counter-term is associated to a smaller absolute value of the renormalized entanglement entropy in comparison to the BK counter-term. We conclude the paper in \S\ref{sec-disc} with a discussion on some intriguing possibilities to explore.   Some complex technical results are presented in Appendix \ref{app-a}.

\section{\boldmath The DBI Counter-Term and Topological \texorpdfstring{{$S^3$}{ }}\  Partition Function}\label{sec-s3pf}

{Since we are interested in the evaluation of the partition function in various contexts,  we will work exclusively in the Euclidean signature.}
We begin with the Euclidean action in four-dimensional asymptotically AdS spacetime with the AdS radius $L$, which we write in terms of four parts, namely,
\baa
I  =  I_{\mathrm{bulk}} + I_{\mathrm{GHY}} + I_{\mathrm{ct, DBI} } + I_{\mathrm{matter}}, \label{itot}
\ea
where
\begin{align}
 I_{\mathrm{bulk}}  &= - \frac{1}{16\pi G} \int\limits_{\mathcal{M}} \dd[4]{x} \sqrt{g} \qty(R  + \frac{6}{L^2}),  \label{ibulk}\\
 I_{\mathrm{GHY}} &=   - \frac{1}{8\pi G} \int\limits_{\partial \mathcal{M}} \dd[3]{x} \sqrt{\gamma} K,  \label{ighy}
\end{align}
with $\gamma_{\mu \nu}$ is the induced metric on the three-dimensional boundary $\partial \mathcal{M}$ of the bulk spacetime $\mathcal{M}$.  The Gibbons-Hawking-York (GHY) term \eqref{ighy} makes the variational principle well-defined \cite{York:1972sj,  Gibbons:1976ue} for the  Dirichlet boundary condition on the metric.  To render the gravitational action finite,  following the standard prescription of holographic renormalization,  one needs to add local counter-terms dependent on the \emph{intrinsic} geometry of the boundary. In this work, we ignore any matter contribution and as stated in \eqref{itot},  we consider the Dirac-Born-Infeld--like (DBI) counter-term \cite{Jatkar:2022zdz},
\baa
I_{\mathrm{ct, DBI} }  = \frac{L^2}{4 \pi G} \int\limits_{\partial \mathcal{M}} \dd[3]{x} \sqrt{-\det B_{\mu \nu} }, \label{idbi}
\ea
where
\baa
B_{\mu \nu}  = R^{(3)}_{\mu \nu} - \frac{1}{2} R^{(3)} \gamma_{\mu \nu} - \frac{1}{L^2}   \gamma_{\mu \nu}  \label{bmndef}
\ea
is defined on the three-dimensional boundary.  {Note that the sign under the square root in \eqref{idbi} is related to how the  tensor $B_{\mu \nu}$ is defined in \eqref{bmndef} --- in the limit of vanishing curvature of the boundary we see that $\sqrt{- \det B_{\mu \nu} } \to L^{-3} \sqrt{\gamma}$, as one would expect. }Given that we are working with this  tensor, the zero-curvature limit essentially determines the sign under the determinant --- it is simply related to the sign of the eigenvalues of the metric tensor. In our set-up,  the metric tensor is positive definite and hence the negativity of the eigenvalues of $B_{\mu \nu}$ persists even after turning on curvature in a weak, perturbative manner.  It is worth emphasizing that there can be situations in which the curvature is large enough so that the eigenvalues change their signature.  While it would be interesting to consider such possibilities,  in the present context,  we focus exclusively on the scenarios in which the eigenvalues have the same sign as in a flat boundary.

However, the holographic counter-term most widely used in the literature is the Balasubramanian-Kraus (BK) counter-term \cite{Balasubramanian:1999re},
\baa
I_{\mathrm{ct, BK} }   = \frac{1}{8\pi G} \int\limits_{\partial \mathcal{M}}  \dd[3]{x} \sqrt{\gamma} \pqty{  \frac{2}{L} + \frac{L}{2} R^{(3)} }.  \label{ibkdef}
\ea
As mentioned previously,  all calculations involving these two counter-terms \eqref{idbi} and \eqref{ibkdef} agree when the boundary $\partial \mathcal{M}$ is pushed to the asymptotic infinity.  The results are, however, different in interesting ways when one considers a finite cut-off. One can,  in principle, construct an infinite number of counter-terms which agree with \eqref{ibkdef} in the asymptotic limit.  The counter-term \eqref{idbi} is rather special among this infinitely many for several reasons.

The DBI counter-term \eqref{idbi} gives rise to a vanishing on-shell action for pure AdS$_4$ with a $S^1 \times S^2$ boundary \cite{Jatkar:2011ue} even for any arbitrary finite radial cut-off.  As explained in detail in \cite{Jatkar:2022zdz},  this feature can be interpreted as the signature of a complete background subtraction in the geometry --  which is also pertinent for describing situations when there are black holes in the bulk.  In \cite{Jatkar:2022zdz},  it was shown that \eqref{idbi} continues to be a good counter-term for describing black hole thermodynamics even in the presence of a Maxwell field and black hole thermodynamics in \emph{any} statistical ensemble in a finite cut-off surface of an \emph{arbitrary} shape.  In fact,  it is possible to consider a notion of \emph{local} thermodynamics in the boundary theory.

In contrast with other possible counter-terms, the on-shell action with the DBI counter-term is also well-behaved \cite{Jatkar:2022zdz} in the flat-space limit,  for which one takes the limit $L \to \infty$.  Therefore,  flat-space thermodynamics \cite{Gibbons:1976ue} can be placed more naturally in the context of holographic renormalization with this counter-term.

\subsection{The 3-Sphere Partition Function at a Finite Radial Cut-Off}\label{subsec-3sphcon}

As mentioned in the introduction,  the negative log of the 3-sphere partition function for a CFT$_3$ is an important quantity, usually referred to as the $F$-function,  which has been argued to decrease monotonically under an RG flow \cite{Jafferis:2011zi}.  This is known as the $F$-theorem in the literature.   Let us first evaluate the 3-sphere partition function holographically by the two different  counter-terms and point out an essential difference between the two.   The bulk {Euclidean} AdS$_4$ metric which is asymptotic to the 3-sphere is given by
\baa
\dd{s}^2 = \dd{\rho}^2 + L^2 \sinh^2 \pqty{ \frac{\rho}{L} } \pqty{ \dd{\theta}^2 + \sin^2 \theta ( \dd{\phi}^2 + \sin^2 \phi \dd{\psi}^2 ) }, \label{ads4metrho}
\ea
where $\rho \geq 0$,  $0 \leq \theta \leq \pi$,  $0 \leq \phi \leq \pi$ and $0 \leq \psi < 2\pi$.  The asymptotic infinity is located at $\rho = \infty$,  which has to be regulated by imposing a finite radial cut-off $\rho  = \rho_C$.  Let us now compute the on-shell action with this radial cut-off.

The bulk action up to the radial cut-off $\rho = \rho_C$ is given by
\baa
I_{\mathrm{bulk}} &= \frac{\pi L^2}{16G}  \bqty{8 - 9\cosh( \frac{\rho_C}{L} ) + \cosh( \frac{3\rho_C}{L} )  }. \label{ibkbulks3}
\ea
The GHY boundary term is given by
\baa
I_{\mathrm{GHY}}=   \frac{3\pi L^2}{16G}  \bqty{\cosh( \frac{\rho_C}{L} ) - \cosh( \frac{3\rho_C}{L} )  }\ . \label{ibkghys3}
\ea
On the other hand, the BK counter-term is given by
\baa
I_{\mathrm{ct, BK} } =  \frac{\pi L^2}{8 G}  \bqty{3\sinh( \frac{\rho_C}{L} ) + \sinh( \frac{3\rho_C}{L} )  }\ . \label{ibkfins3}
\ea
The total on-shell action, sum of the three previous expressions,  is the candidate $F$-function corresponding to the BK renormalization scheme,
\baa
F_{\mathrm{BK}} (\rho_C) =  \frac{\pi L^2}{2G} \pqty{ 1 - \frac{3}{4} e^{- \rho_C/L } - \frac{1}{4} e^{- 3\rho_C/L }   }. \label{fbkfin}
\ea
As $\rho_C \to \infty$,  we have $F =  \pi L^2 / 2 G$ as one would have expected \cite{Nishioka:2018khk}.   We note however the interesting feature that for any finite value of the cut-off $\rho_C$,
\baa
F_{\mathrm{BK} } \qty(\text{finite } \rho_C ) < F_{\mathrm{BK} } (\infty).
\label{fbkmon}
\ea
This implies that the BK scheme,  taken in itself,  is consistent with the monotonicity property of the $F$-function, now in the context of finite cut-off holography.   It signifies that the degrees of freedom along the RG flow induced by the BK counter-term \eqref{ibkdef} decrease in number.

However,  if one augments the BK counter-term with terms which vanish in the asymptotic limit,  we would not,  in general,  expect such monotonicity properties to hold.  Since the monotonicity property --- which need not necessarily refer to a \emph{strict} monotonicity as in the example above --- can be taken as a fundamental property  of RG flows,  it can help establish the consistency of the different candidate counter-terms associated with the RG flow.

Let us now see how the result changes when we consider the DBI counter-term instead.  We find that
\baa
I_{\mathrm{ct,  DBI} }   =  \frac{\pi L^2}{8 G} \bqty{3\cosh( \frac{\rho_C}{L} ) + \cosh( \frac{3\rho_C}{L} )  }. \label{icts3dbi}
\ea
All the cosh terms from the bulk \eqref{ibkbulks3} and the GHY term \eqref{ibkghys3} now cancel exactly and we have for any $\rho_C$,
\baa
F_{\mathrm{DBI}} (\rho_C) = \frac{\pi L^2}{2G}.  \label{fdbi}
\ea
This is independent of $\rho_C$.  This cut-off independence is crucial to the scenario in \cite{Jatkar:2011ue, Jatkar:2022zdz} and analogous to the feature that for the $S^1 \times S^2$ boundary that  the renormalized on-shell action for pure AdS$_4$ vanishes.  This can also be interpreted as performing some sort of background subtraction \cite{Jatkar:2022zdz}.  This remarkable feature immediately raises a question which we address in the next subsection.

\subsection[Is \texorpdfstring{$F_{\rm DBI}$}{FDBI}\ a Topological Invariant?]{Is {\boldmath $F_{\text{DBI}}$  a Topological Invariant?}}\label{subsec-fdbitop}

Since the 3-sphere partition function is an invariant  under the change of the radius of the spherical cut-off surface,  this immediately suggests to us the possibility that the partition function itself does not depend on  arbitrary smooth deformations of the cut-off 3-sphere.  In other words,  it could be shape-independent or a topological invariant.

We have to choose some cut-off surface that explicitly breaks the spherical symmetry.  We could consider linear perturbations around a spherical surface $\rho = \rho_C$ as was done for black hole thermodynamics \cite{Jatkar:2022zdz}. Even at the linear level,  obtaining thermodynamics to work out was rather non-trivial.  In the case at hand,  however,  let us work at the full non-linear level.   It is convenient to work with a slightly different radial coordinate than before,
\baa
r = L \sinh (\rho/L),  \label{newrad3sp}
\ea in which the answer is relatively simple to write down.  In this coordinate, the $\mathrm{AdS}_4$ metric with $S^3$ boundary is given by
\baa
\dd{s}^2 = \frac{\dd{r}^2}{1 + \frac{r^2}{L^2} } + r^2 \pqty{ \dd{\theta}^2 + \sin^2 \theta ( \dd{\phi}^2 + \sin^2 \phi \dd{\psi}^2 ) } \,   \label{ads4rcos3}.
\ea

In this coordinate system,  let the cut-off surface be located at
\baa
r=  L\, Y(\theta), \label{3spcutoff}
\ea
where $Y$ is an arbitrary single-valued function  of only the first angular coordinate $\theta$.  The ``origin'' $r = 0$ is assumed to be enclosed by the surface. We will see while this choice is relatively simple, it leads to rather complex expressions.
The induced metric on the 3-dimensional cut-off surface is thus given by
\baa
\dd{s}^2 = L^2 \bqty{ \pqty{\frac{Y'(\theta)^2}{1+ Y(\theta)^2} + Y(\theta)^2} \dd{\theta}^2 + Y(\theta)^2  \sin^2 \theta ( \dd{\phi}^2 + \sin^2 \phi \dd{\psi}^2 ) }.  \label{induced3sph}
\ea
Since the surface {under consideration is taken to be smooth and topologically equivalent to a 3-sphere,}  we allow for any arbitrary (twice differentiable) function $Y$ which has the following smoothness conditions\footnote{{One can also consider a situation in which there are conical defects at the poles.  The final result will then be different from the case discussed in this article.  One might need to consider  corner contributions such as the  Hayward term \cite{Takayanagi:2019tvn} for such geometries.  It would be interesting to pursue this line of investigation further. }} at the poles:
\baa
Y' (\theta \to 0, \pi)  = 0. \label{ysmo}
\ea

We write the answer for the on-shell action in the form of an integral over the angular coordinate $\theta$ (the bulk term involves a radial integral from $r = 0$ up to $r = L \, Y(\theta)$ and the $\phi$ and $\psi$ integrals are performed for each of the terms):
\baa
I_{\mathrm{total}} =  - \frac{L^2}{16\pi G} \int_0^{\pi} \dd{\theta} \bqty{  {\cal I}_{\rm bulk}  (\theta) +  {\cal I}_{\rm GHY}  (\theta) +  {\cal I}_{\rm DBI}  (\theta)  },   \label{itotals3}
\ea
where the subscripts have obvious meanings.  We find from explicit calculations that
\baa
{\cal I}_{\rm bulk}  (\theta)  = -8 \pi  \sin ^2 \theta  \qty( Y(\theta )^2 \sqrt{Y(\theta )^2+1} -2 \sqrt{Y(\theta )^2+1}+2),  \label{ibulks3}
\ea
and
\baa
{\cal I}_{\rm GHY}  (\theta) &= \frac{8\pi Y \sin \theta  }{\sqrt{Y^2+1} (Y'^2+Y^4+Y^2) } \Bigg[ Y^2 (Y^2+1)   \pqty{3 Y (Y^2+1)-Y''} \sin \theta \\
&\quad +Y (5 Y^2+4) Y'^2  \sin \theta -2  Y'^3 \cos \theta  -2 Y^2 (Y^2+1)  Y' \cos \theta \Bigg].  \label{ighys3}
\ea
(The explicit $\theta$-dependence of $Y$ has been suppressed in the previous expression for brevity.)
Finally, we find the following expression for the DBI counter-term:
\baa
{\cal I}_{\rm DBI}  (\theta) &= - \frac{16\pi}{\sqrt{Y^2 +1}} \pqty{  \frac{(Y^2+1) Y \sin \theta -Y' \cos \theta}{Y^4+Y^2+Y'^2} }^2 \times\\
&\quad \bqty{2 Y'^2 + Y^2 \pqty{ (1 + Y^2)^2 + 3 Y'^2 }  - Y (1+ Y^2) Y''  }. \label{idbis3}
\ea

It is perhaps obvious to the reader from the foregoing expressions that the sum of the three functions \eqref{ibulks3},  \eqref{ighys3} and \eqref{idbis3} is a very complicated non-linear function of the cut-off surface $Y(\theta)$ and its derivatives $Y'(\theta)$ and $Y''(\theta)$.   It is nevertheless possible to write  this complicated total integrand in the form
\baa
 {\cal I}_{\rm bulk}  (\theta) +  {\cal I}_{\rm GHY}  (\theta) +  {\cal I}_{\rm DBI}  (\theta)  = \dv{H(\theta)}{\theta}  - 16\pi \sin^2 \theta,  \label{totints3}
\ea
where
\baa
H(\theta) &= 8\pi \tan^{-1} \frac{Y'(\theta )}{Y(\theta ) \sqrt{Y(\theta )^2+1} }  - 8 \pi \sqrt{Y(\theta )^2+1} \sin 2 \theta  \\
&\quad + \frac{8 \pi Y(\theta ) \sqrt{Y(\theta )^2+1} \bqty{ Y'(\theta ) \left( Y(\theta )^2 \sin ^2 \theta-\cos 2 \theta \right)+ Y(\theta) (Y(\theta )^2+1 ) \sin 2 \theta }}{Y'(\theta )^2+Y(\theta )^4+Y(\theta )^2}.
\ea
On account of the smoothness property \eqref{ysmo},  we immediately see that $H(\theta) \to 0$ near the poles $\theta  = 0, \pi$.  Therefore, on performing the integration $\theta$ in \eqref{itotals3},  the dependence on the shape of the cut-off surface $Y(\theta)$ disappears completely.

The integral of the remaining term in \eqref{totints3} gives precisely the constant answer \eqref{fdbi} for the spherical cut-off surface.  Thus, we conclude that the DBI counter-term indeed has some topological property --- it is extremely likely that this feature would hold true for more general cut-off surfaces.  On the other hand,  it is not possible for the BK term to exhibit a similar property simply because of its explicit dependence on the cut-off radius.

How do we place this result in the context of RG flows? Since a constant radial cut-off corresponds to a UV cut-off in the boundary theory,  the present set-up immediately invites the interpretation that the cut-out surface corresponds to a \emph{position-dependent} UV cut-off in the field theory.  It would be interesting to make this intuition more precise on the field theory side.

\section{Ryu-Takayanagi Surfaces and the DBI Counter-Term}\label{sec-rtdbi}

The results in the previous section point towards quite remarkable aspects of the DBI counter-term \eqref{idbi}.   It is therefore interesting to consider other quantities in the setting of finite cut-off holography and examine if such remarkable properties extend to other scenarios.  One of the most interesting quantities to consider in this context is holographic entanglement entropy,   obtained through the Ryu-Takayanagi (RT) formula \cite{Ryu:2006bv,  Ryu:2006ef}.  Let us consider the CFT dual of a static asymptotically AdS spacetime.  The entanglement entropy corresponding to a spatial boundary region $\mathcal{R}$ on the field theory side is then given by
\baa
S_{\mathrm{EE}} (\mathcal{R}) = \min \frac{\text{Area}(\sigma_{\partial \mathcal{R}} )}{4 G}, \label{rt-form}
\ea
where $\sigma_{\partial \mathcal{R}}$ is a bulk codimension-2 spacelike surface which is anchored to the entangling surface $\partial \mathcal{R}$ on the boundary.  The RT formula \eqref{rt-form} says that the entanglement entropy is given by the minimal area of such a surface --- it is remarkable in that it reduces the calculation of entanglement entropy in a strongly coupled field theory to a geometric problem.

The formula \eqref{rt-form} taken as such yields a divergent answer as the bulk cut-off is taken to asymptotic infinity.    In this  limit,  as usual,  the IR divergence from the boundary of the AdS spacetime is interpreted as a UV divergence on the field theory side.   The divergence is exactly of the correct form to explain the area-law divergence in the entanglement entropy in field theory \cite{Srednicki:1993im}.   In general dimensions,  the leading area-law divergence is followed by sub-leading divergences \cite{Ryu:2006bv,  Ryu:2006ef}.  The goal of the present section is to perform calculations similar in spirit to those in \S\ref{sec-s3pf} and \cite{Jatkar:2022zdz} to examine the effects of different gravitational counter-terms,  \eqref{ibkdef} and \eqref{idbi}.
However,  the formula  \eqref{rt-form} as stated has no bearing  on any boundary counter-terms.  One method to incorporate their effects,  which has sometimes been employed in the literature, is as follows: one calculates the on-shell action (including the boundary terms) for the spacetime region  between the RT surface and the boundary.  This is said to give rise to the ``entanglement free energy'' $\beta_{\mathrm{ent}} F_{\mathrm{ent}}$, where $\beta_{\mathrm{ent}}$ is the ``inverse entanglement temperature'',  from which one extracts a renormalized value of the entanglement entropy.  While the counter-terms play a prominent role in this calculation,  it does not yield physically reasonable answers.

One is thus led to look for another more direct method of evaluating the entanglement entropy with the counter-terms.  One needs to follow the steps in the derivation of the holographic entanglement entropy formula through the replica method on the gravitational path integral \cite{Lewkowycz:2013nqa}, but with a \emph{renormalized} gravitational action.
The general method of obtaining a \emph{renormalized entanglement entropy} has already been outlined in \cite{Taylor:2016aoi}.  One can extract the finite part of the entanglement entropy by using results pertaining to the behavior of curvature invariants on conical spaces. Let $I (n)$ be the ``bare'' gravitational action (bulk action + the boundary GHY term) corresponding to an $n$-fold replica in $d+1$-dimensional spacetime.  It was shown in \cite{Lewkowycz:2013nqa} that
\begin{equation}
S_{\mathrm{EE}} = n \pdv{n} \eval{ \pqty{  I(n) - n I (1) } }_{n=1} =  \min \bqty{ \frac{1}{4G} \int \dd[d-1]{x} \sqrt{h} },  \label{seelm}
\end{equation}
where $h$ is the induced metric on the co-dimension two bulk surface (satisfying the usual conditions).

The authors of \cite{Taylor:2016aoi} had used, in addition to the above term,  the boundary counter-term action to define the renormalized  entropy.   The boundary terms at asymptotic infinity were argued to have a vanishing contribution to the entropy  in \cite{Lewkowycz:2013nqa}, but the analysis assumed the standard GHY boundary term and did not include the curvature-dependent counter-terms on the boundary.   The key insight of \cite{Taylor:2016aoi} was that boundary counter-terms made out of boundary curvature invariants \emph{do} give rise to a non-vanishing (actually divergent) contribution and this is precisely what gives a finite, renormalized value of the entropy.   This is related to the observation that near $n=1$,  the intrinsic curvatures of the replicated boundary manifold have the following form \cite{Fursaev:2013fta}
\baa
R^{(n)} &= R + 4\pi (1 -n) \delta_{\partial \mathcal{R}} + \mathcal{O} (n-1)^2,  \\
R^{(n)}_{\mu \nu} &= R_{\mu \nu} + 2\pi (1 -n) n_\mu n_\nu \delta_{\partial \mathcal{R}} + \mathcal{O} (n-1)^2, \label{sqcones}
\ea
where $ \delta_{\partial \mathcal{R}} $ is the delta-function which localizes any boundary integral to the co-dimension 2 entangling surface and $n_\mu n_\nu = \sum_a n^a_\mu n^a_\nu$,  where $n^a_\mu$ is the unit normal vector corresponding to one of the two directions $a$.  {One of the two directions labeled by $a$ is the temporal direction (since the entropy is being calculated on a constant time-slice) and the other direction is associated with the entangling surface on the constant time-slice. }  {To be more precise,  near the conical singularity of the replica manifold,  it is possible to choose coordinates so that the metric in the vicinity of the tip of the cone  $(r= 0)$ looks like,
\begin{equation}
\dd{s}^2 = \dd{r}^2 + n^2 r^2 \dd{\theta}^2 + \dd{s}_\perp^2,
\end{equation}
where  $\dd{s}_\perp^2$ is the metric in the transverse direction and $\theta \sim \theta + 2\pi$.  Slightly away from the tip of the cone,  the geometry locally looks like a flat 2-plane tensored with the transverse space.   The additional curvature contribution due the delta function is important only when an integration is performed. } The use of the superscript in \eqref{sqcones} is to be distinguished from eq. \eqref{idbi}.

\subsection{A Flat Boundary}\label{subsec-rtflat}

For a CFT on $\mathbb{R}^{1,3}$, the intrinsic curvature $R_{\mu \nu}$ vanishes --- nevertheless,  one still obtains a non-trivial boundary contribution to the RT area functional because of the foregoing result. The authors of \cite{Taylor:2016aoi} explicitly obtained this contribution due to the BK counter-term \eqref{ibkdef} for such a flat boundary.   The unreplicated on-shell gravitational action $I$ is identical for the BK and DBI counter-term since the boundary is flat.  It is not \emph{a priori} obvious that the entanglement entropy would agree.  Let us consider a replica variation of the DBI counter-term \eqref{idbi},
\baa
\delta I_{\mathrm{ct}}  = \frac{L^2}{8\pi G} \int \dd[3]{x} \sqrt{- \det B} \pqty {B^{-1} }^{\mu \nu} \delta B_{\mu \nu} \label{dict}.
\ea
Given the form of the tensor \eqref{bmndef} and the replica relations \eqref{sqcones} about the boundary intrinsic curvature, we find
\baa
\delta B_{\mu \nu}  =  2\pi (1-n) \pqty{ n_\mu n_\nu -  \gamma_{\mu \nu} } \delta_{\partial \mathcal{R}}. \label{dbmng1}
\ea
The unperturbed $B_{\mu \nu}$ for a flat boundary is given  by \eqref{bmndef}
\baa
B_{\mu \nu} = - \frac{1}{L^2} \gamma_{\mu \nu}.  \label{bmnflat}
\ea
It is thus easy to show that
\baa
\delta I_{\mathrm{ct}}  = \frac{L}{4G} (1-n) \int \dd{x}  \sqrt{ \gamma_{\partial \mathcal{R}} }, \label{dict1n}
\ea
where $\gamma_{\partial \mathcal{R}}$ is the pull-back of the induced metric (i.e.,  pull-back of a pull-back) on the entangling surface.

Thus, the boundary counter-term to the RT area term is given by
\baa
\eval{\pdv{(\delta I_{\mathrm{ct}}) }{n} }_{n=1} = -\frac{L}{4G} \int \dd{x}  \sqrt{\gamma_{\partial \mathcal{R}} } . \label{ictdn}
\ea
This is also the same quantity obtained by \cite{Taylor:2016aoi} for the BK counter-term.  One has to add  the bulk area piece \eqref{rt-form} with the boundary counter-term \eqref{ictdn} to obtain the total renormalized entropy.   It is encouraging that the answers for the entanglement entropy agree for the two counter-terms with a planar boundary.

An appealing feature of the method of \cite{Taylor:2016aoi} is that it is applicable for boundary regions of any arbitrary shape.  It can also be directly applied in the context of finite cut-off holography,  including situations where the cut-off surface has an irregular shape.

Before we move on to a description of situations in which the DBI term yields a different answer from the BK term,  let us discuss an example,  previously considered in \cite{Taylor:2016aoi}, in the context of holographic renormalization.   This example concerns a circular entangling surface of radius $\rho_0$ on the boundary theory.  The Euclidean AdS$_4$ metric well-suited to describe this situation is given by
\baa
\dd{s}^2 = \frac{L^2}{z^2} \pqty{ \dd{t}^2 + \dd{z}^2 + \dd{\rho}^2 + \rho^2 \dd{\phi} ^2 }.  \label{adspoin}
\ea
In this coordinate system,  the minimal RT surface is part of a sphere
\baa
\rho(z) = \sqrt{z_*^2 - z^2},  \label{rtsurf1}
\ea
where $z_*$ is the turning point in the bulk defined by $z_* = \sqrt{\rho_0^2 + \epsilon^2}$,  where the boundary cut-off surface is at $z = \epsilon$.  We can thus calculate the bulk piece \eqref{rt-form} in the entanglement entropy
\baa
S_{\mathrm{bulk}} = \frac{\pi L^2 z^*}{2G} \pqty{ \frac{1}{\epsilon}  - \frac{1}{z^*} }.  \label{sbulks1}
\ea
The pull-back of the boundary metric on the entangling surface  $\rho = \rho_0$ at a constant time-slice is given by
\baa
\dd{s}^2 = \frac{L^2 \rho_0^2}{\epsilon^2} \dd{\phi}^2.  \label{induced1}
\ea
Thus,  from \eqref{ictdn}, we find the total renormalized entropy at the cut-off $\epsilon$,
\baa
S_{\mathrm{REE}} (\epsilon) &= -\frac{\pi  L^2}{2 G}  \bqty{ 1  - \frac{\sqrt{\rho_0^2 + \e^2} - \rho_0}{\epsilon} }. \label{sreedisc1}
\ea
In the limit $\epsilon \to 0$,  we find the known result $- \pi L^2/2 G$ for the finite piece in entanglement entropy.  That this quantity in the $\epsilon \to 0$ limit is the negative of the asymptotic value of the $F$-function \eqref{fdbi} is not an accident,  see \cite{Casini:2011kv}. The negativity of the renormalized entropy is also another expected feature.  In the context of a finite cut-off, we notice that
\baa
| S_{\mathrm{REE}} (\epsilon > 0) | < | S_{\mathrm{REE}} (0) |  \label{entmon}.
\ea
This is consistent with the expectation that as physical degrees of freedom are integrated out,  the absolute value of the renormalized entropy decreases. The decrease here is strictly monotonic irrespective of whether one uses the BK or DBI counter-terms.  Furthermore,  while the $F$-function is a constant for the DBI counter-term for a varying cut-off of any shape,  the disk entanglement entropy is not.

\subsection{Renormalized Entropy in Global AdS Spacetime}\label{subsec-rtglobcon}

It is reassuring to find agreement for the values of the renormalized entanglement entropy for the two counter-terms on a flat boundary.  Let us now examine a situation in which the counter-terms \eqref{idbi} and \eqref{ibkdef} themselves would not agree at a finite radial cut-off.  To this end, we consider the Euclidean section of global AdS spacetime,  with the  asymptotic boundary $S^1 \times S^2$.   The results in \S\ref{sec-s3pf} and \cite{Jatkar:2011ue,  Jatkar:2022zdz} suggest that there might be some manifestation of cut-off independence for the entanglement entropy as well.   In the case of an $S^2$ spatial boundary,  the most obvious region to consider is an azimuthally symmetric ``polar cap'',  which in the standard angular coordinates is given by $\theta < \theta_0$.  Unlike in the case of the disk on a flat boundary,  the RT surface corresponding to this very symmetric entangling region is not known in a simple closed form.  For a general value of $\theta_0$, one has to solve the RT surface numerically,  as we shall do in due course.  However,  for the present purpose,  we note that there is a special region for which the RT surface is known analytically: the hemisphere.

The line element for the Euclidean section of the global AdS spacetime is given by
\baa
\dd{s}^2 =  \pqty{1 + \frac{r^2}{L^2} } \dd{t}^2 + \frac{\dd{r}^2}{ \pqty{1 + \frac{r^2}{L^2} }} + r^2 \pqty{ \dd{\theta}^2 + \sin^2 \theta \dd{\phi}^2 }.  \label{adsmet}
\ea
For the hemisphere,  the boundary entangling surface is given by the great circle $\theta = \pi/2$.  The bulk RT surface for a spherical cut-off is given by the ``equatorial disk'' $\theta = \pi/2$,  as can also be checked by inserting this in the Euler-Lagrange equations for the area functional.   We now proceed to calculate $S_{\mathrm{REE, BK}}$ and $S_{\mathrm{REE, DBI}}$ one after another and examine their cut-off dependence.

The regulated area of the RT surface is given by
\baa
A = 2\pi L \qty( \sqrt{r_C^2 + L^2} - L).   \label{eqdisreg1}
\ea
The BK counter-term  \eqref{ibkdef} is linear in the boundary Ricci scalar,  and therefore under the replica variation \eqref{sqcones}, the boundary counter-term for the entanglement has the same form as \eqref{ictdn} for any boundary and is curvature-independent,
\baa
 S_{\mathrm{ct, BK}} &=  -\frac{L}{4G} \int \dd{x}  \sqrt{\det \gamma_{\partial \mathcal{R}} } = - \frac{\pi L r_C}{2G}  \label{sctbk2}.
\ea
Therefore,  the renormalized entanglement entropy from the BK term is
\baa
S_{\mathrm{REE, BK}} = \frac{\pi L}{2G} \pqty{ \sqrt{r_C^2 + L^2} - L - r_C  },  \label{sreebkeq1}
\ea
and it depends explicitly on the radial cut-off.  In the limit the cut-off is taken to asymptotic infinity $r_C \to \infty$,
\baa
S_{\mathrm{REE}} = - \frac{\pi L^2}{2G}.  \label{treebkeq1}
\ea
This value is identical to the renormalized disk entanglement entropy \eqref{sreedisc1} in the asymptotic limit and the negative of the value of the $F$-function.  In fact,  the value  \eqref{treebkeq1} is not special to the equatorial entangling surface and has the same value for any polar cap angle $\theta_0$ \cite{Casini:2011kv}.

Let us now carry out the analogous computation for the DBI case.   For the sake of generality,  we consider a static spherically symmetric metric
\baa
\dd{s}^2 =  f(r) \dd{t}^2 + \frac{\dd{r}^2}{g(r)} + r^2 \pqty{ \dd{\theta}^2 + \sin^2 \theta \dd{\phi}^2 },
\ea
and the polar cap region, $\theta < \theta_0 $.
The general variation of the DBI action was presented in eqs.  \eqref{dict} and \eqref{dbmng1}.   For the flat case,  the $B_{\mu \nu}$ tensor was trivial \eqref{bmnflat},  but it is non-trivial in the present context because of the curved boundary geometry.  From the induced metric on the spherical cut-off surface $r = r_C$,
\baa
\dd{s}^2 = \gamma_{\mu \nu} \dd{x^\mu} \dd{x^\nu} = f(r_C) \dd{t}^2 + r_C^2 \pqty{ \dd{\theta}^2 + \sin^2 \theta \dd{\phi}^2 },
\ea
we can easily find the non-vanishing components of the boundary Ricci tensor and hence the Ricci scalar,
\baa
R_{\theta \theta} = \frac{R_{\phi \phi}}{\sin^2 \theta} = 1, \qquad R = \frac{2}{r_C^2}.  \label{bricci1}
\ea
This yields the components of the $B$ tensor to be,
\baa
B_{\mu \nu} = - \mathrm{diag} \pqty{ \pqty{ \frac{1}{L^2} +\frac{1}{r_C^2} } f(r_C) ,  \frac{r_C^2}{L^2},  \frac{r_C^2}{L^2} \sin^2 \theta  }. \label{bmntensors2}
\ea
The relation involving the two normal vectors for the entangling surface
\baa
 n_\mu n_\nu -  \gamma_{\mu \nu}  =  \mathrm{diag} \pqty{ 0,  0,  -r_C^2 \sin^2 \theta  }, \label{nmnn1}
\ea
 immediately leads to
\baa
\pqty {B^{-1} }^{\mu \nu} \delta B_{\mu \nu}  = 2\pi (1-n) L^2 \delta_{\partial \mathcal{R}}
&=  2\pi (1-n) L^2 \frac{1}{r_C \sqrt{f(r_C) }} \delta(t-t_0) \delta(\theta - \theta_0),  \label{deltaspheric}
\ea
with the appropriate normalization for the delta functions. The entropy is evaluated at the constant time slice $t =  t_0$. We thus find the entanglement counter-term at any radius and polar cap angle $\theta_0$ to be
\baa
 S_{\mathrm{ct, DBI}} =  - \frac{\pi  L \sqrt{L^2 + r_C^2} \sin \theta_0}{2G}.  \label{sctdbipol1}
\ea
Therefore,  for the equatorial entangling surface with $\theta_0 = \pi/2$,  we find, using the bulk area term \eqref{eqdisreg1},
\baa
S_{\mathrm{REE, DBI}} &=  \frac{A}{4G} +  S_{\mathrm{ct, DBI}} = - \frac{\pi L^2}{2G}.  \label{sreedbisph1}
\ea
This is a pleasing result --- the renormalized entanglement entropy for the DBI counter-term is independent of the radial cut-off. This is completely consistent with the intuition spelt out in the beginning of this subsection.  If we consider the positive quantity $\qty| S_{\mathrm{REE} } (r_C) |$,  we see the same sort of monotonicity property for the BK and DBI counter-terms as for the $F$-function in \S\ref{sec-s3pf}.  The foregoing result, however,  concerns the rather special equatorial RT surface for which an analytical solution is available.  It remains to be seen whether such special properties extend to polar cap regions as well. Before we investigate that question numerically, let us consider another important question that can be dealt with analytically --- whether the entanglement entropy for the analog of the great circle entangling region depends on the shape of the cut-off surface.

\subsection{Arbitrary Cut-Out Equatorial Disk}\label{subsec-rtcut}

\begin{figure}
\centering
\includegraphics[scale=0.6]{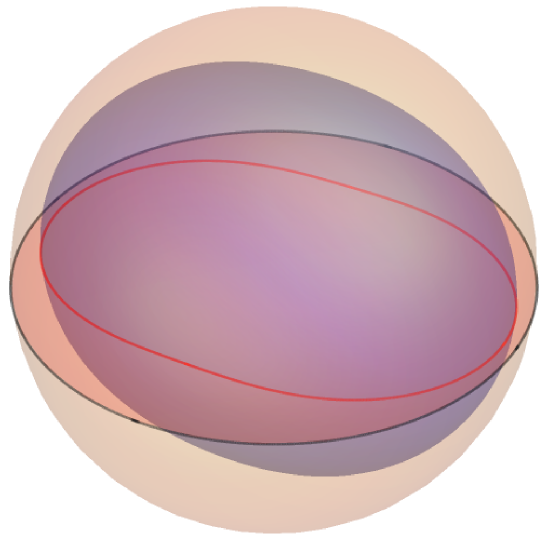}
\caption{An illustration of the set-up under consideration in \S\ref{subsec-rtcut}. The sphere represents the asymptotic infinity. The surface in blue is a non-spherical surface at finite cut-off, which intersects the equatorial Ryu-Takayanagi surface along the red curve.}
\label{fig-cutout}
\end{figure}

Let us consider the following time-independent cut-off surface,
\baa
r = L f(\theta, \phi) \label{defsig}
\ea
in the global AdS spacetime \eqref{adsmet}. Here $f(\theta, \phi)$ is a single-valued function and the surface encloses the origin. The scenario is depicted in Figure \ref{fig-cutout}.

Based on the results in \S\ref{subsec-fdbitop} and \S\ref{subsec-rtglobcon},  It is then tempting to conjecture that the REE for the $\theta = \pi/2$ entangling surface on the boundary is the same for any arbitrary twice-differentiable function $f(\theta, \phi)$,
\baa
S_{\mathrm{REE, DBI}} =  - \frac{\pi L^2}{2G}. \label{sree}
\ea
It is somewhat of a surprise that the strongest form of the above conjecture is not true.  Nevertheless,  it turns out that the conjecture is true for a large class of surfaces, which we will specify soon.

Since we are required to consider no further than the second derivative of the function \eqref{defsig} on the entangling surface $\theta = \pi/2$,  we need no more than three independent functions of $\phi$,
\baa
\eval{f ( \theta , \phi )}_{ \theta = \frac{\pi}{2} } &\equiv X(\phi),  &\quad \eval{\pdv{f ( \theta , \phi )}{\phi}}_{ \theta = \frac{\pi}{2} } &= X'(\phi), \quad \eval{\pdv[2]{f ( \theta , \phi )}{\phi}}_{ \theta = \frac{\pi}{2} } &= X''(\phi), \\
\eval{\pdv{f ( \theta , \phi )}{\theta}}_{ \theta = \frac{\pi}{2} } &\equiv Y(\phi), &\quad \eval{\pdv{f ( \theta , \phi )}{\theta}{\phi}}_{ \theta = \frac{\pi}{2} } &= Y'(\phi), \\
\eval{\pdv[2]{f ( \theta , \phi )}{\theta}}_{ \theta = \frac{\pi}{2} } &\equiv Z(\phi). \label{fdefs}
\ea
All the functions involved must be $2\pi$-periodic in $\phi$,  $\{ X, Y, Z \} (\phi + 2\pi )  = \{ X, Y, Z \} (\phi )$.

As noted previously,  the bulk Ryu-Takayanagi surface in global AdS$_4$ anchored to the $\theta = \pi/2$ curve is just the disk $\theta = \pi/2$  in bulk spacetime.  We must, however, be careful when the azimuthal symmetry is broken by the boundary conditions.  Using the metric \eqref{adsmet} the area functional for the general case involving a dependence on $\phi$ can be written as,
\baa
A &=  \int \dd{r} \dd{\phi} \frac{r}{\sqrt{1+ \frac{r^2}{L^2} }} \sqrt{ \pqty{ \pdv{\theta}{\phi} }^2 + \sin^2 \theta \bqty{1 + r^2 \qty(1+\frac{r^2}{L^2} )  \pqty{\pdv{\theta}{r} }^2}  },
 \label{ara1}
\ea
where $\theta = \theta (r, \phi)$ is taken as the dependent variable.  It is easy to see that the bulk equation of motion obtained by varying this functional is still solved by the constant function $\theta (r,\phi) = \pi/2$.  Therefore, the RT surface in this case continues to be the equatorial plane.   We can thus obtain the bulk piece easily:
\baa
\frac{A}{4G} =  - \frac{\pi L^2}{2G} + \frac{L^2}{4G} \int_0^{2\pi} \dd{\phi} \sqrt{X(\phi )^2+1} .  \label{area2}
\ea

The induced metric on the surface defined by \eqref{defsig} is given by
\baa
\frac{\dd{s}^2}{L^2} &= \pqty{1 + f^2} \frac{\dd{t}^2}{L^2} +  \bqty{f^2 + \frac{1} {1 + f^2} \pqty{\pdv{f}{\theta}}^2 } \dd{\theta}^2  \\
&\quad + \bqty{f^2 \sin^2 \theta + \frac{1} {1 + f^2} \pqty{\pdv{f}{\phi}}^2 } \dd{\phi}^2  + \frac{2}{1+ f^2} \pdv{f}{\theta} \pdv{f}{\phi} \dd{\theta} \dd{\phi}. \label{dfinduced}
\ea

The components of the $B$-tensor are quite complicated.  The non-vanishing components of the Ricci tensor for the metric \eqref{dfinduced} are given in Appendix \ref{app-a}, from which one can determine the components of the $B$-tensor. The determinant for the $B$-tensor reads
\baa
\sqrt{- \det B} &= \frac{L^{-1} X^2}{\left(X^4+X^2+\left(X'\right)^2+Y^2\right)^2} \Bigg[  X^9+3 X^7-X^6 \left(X''+Z\right)-2 X^4 \left(X''+Z\right)\\
&\quad -2 \left(Z (X')^2+Y (Y X''-2 X' Y')\right) +X \left(Z X''+2 \left(X'\right)^2+2 Y^2-\left(Y'\right)^2\right)\\
&\quad -X^2 \left((3 Y^2+1) X''-6 Y X' Y'+3 Z \left(X'\right)^2+Z\right)  \\
&\quad  +3 X^5 \left(\left(X'\right)^2+Y^2+1\right)+X^3 \left(Z X''+5 \left(X'\right)^2+5 Y^2-\left(Y'\right)^2+1\right)  \Bigg].  \label{detbtensor}
\ea
On the other hand,  the components of $n_\mu n_\nu - \gamma_{\mu \nu}$ on the entangling surface are given by
\baa
n_\mu n_\nu - \gamma_{\mu \nu} = L^2
\begin{pmatrix}
 0 & 0 & 0 \\
 0 & -\frac{Y^2 \left(X'\right)^2}{\left(X^2+1\right) \left(X^4+X^2+\left(X'\right)^2\right)} & -\frac{Y X'}{X^2+1} \\
 0 & -\frac{Y X'}{X^2+1} & -X^2-\frac{\left(X'\right)^2}{X^2+1}
\end{pmatrix} , \label{nmunueq}
\ea
where the components are ordered as $(t,\theta, \phi)$.

At the end,  we find the remarkably compact result,
\baa
&\sqrt{- \det B } (B^{-1})^{\mu \nu} (n_\mu n_\nu - \gamma_{\mu \nu}) \\
&= \frac{L X \left(X^6+2 X^4-X \left( 1+ X^2 \right) X''+2 \left(X'\right)^2+X^2 \left(3 \left(X'\right)^2+1\right)\right)}{X^4+X^2+\left(X'\right)^2}  \label{sqdetb}
\ea
which, the reader will notice, is also independent of the functions $Y$ and $Z$.

However,  we also have to include appropriate normalization of the delta function localizing to the boundary. We see that we must have
\baa
\delta_{\partial \mathcal{R}} = \frac{1}{L} \sqrt{\frac{X^4+X^2+\left(X'\right)^2}{\left(X^4+X^2\right) \left(X^4+X^2+\left(X'\right)^2+Y^2\right)}} \delta ( t- t_0) \delta\pqty{ \theta - \frac{\pi}{2} }.  \label{deltafnc}
\ea
Note that we have already used $\theta = \pi/2$ in simplifying all the foregoing expressions.

Thus,  the total renormalized entanglement entropy is
\baa
S_{\rm REE } &=  -\frac{L^2}{4G} \int_0^{2\pi} \dd{\phi} \Bigg[ \frac{X^6+2 X^4-X  \left( 1+ X^2 \right) X''+2 \left(X'\right)^2+X^2 \left(3 \left(X'\right)^2+1\right)}{\sqrt{(X^2 + 1)  \left(X^4+X^2+\left(X'\right)^2+Y^2\right)\left(X^4+X^2+\left(X'\right)^2\right)}} \\
&\quad -  \sqrt{X^2 +1}  \Bigg] - \frac{\pi L^2}{2G}.
\ea

In order to argue for the shape-independence of $S_{\rm REE } $,  we want to show that the $\phi$-integrand is a total derivative.  However,  we immediately see that it is not possible because the free function $Y(\phi)$ appears in the integrand,  but its derivative does not.  Therefore, for the most general cut-off surface, it is not possible to argue that the REE is independent of the cut-off shape. So the strongest form of the conjecture \eqref{sree} is  not correct.

How about the case when $Y(\phi) = 0$? In this case, the integrand is
\baa
\frac{\left(2 X(\phi )^2+1\right) X'(\phi )^2-X(\phi ) \left(X(\phi )^2+1\right) X''(\phi )}{\sqrt{X(\phi )^2+1} \left(X'(\phi )^2+X(\phi )^4+X(\phi )^2\right)} =- \dv{\phi} \tan^{-1} \frac{X'(\phi)}{X(\phi) \sqrt{X(\phi)^2 + 1} }, \label{sreetotsads}
\ea
which is  a total derivative! Since $X$ is a periodic function of $\phi$, the renormalized entanglement entropy is
\baa
S_{\rm REE }  = - \frac{\pi L^2}{2G}.  \label{sreedisc}
\ea
Thus, we see that the conjecture that the REE for the equatorial disk for an arbitrary cut-off surface being an invariant is indeed true in the case $Y(\phi) = 0$,  i.e,  the situation in which the equatorial RT surface intersects the cut-off surface orthogonally.   We know that the Ryu-Takayanagi surface always touches the asymptotic boundary orthogonally,  but it need no longer be the case for a finite cut-off surface. It is also not difficult to see that if we choose a constant value of the function $X(\phi)$,  any non-zero $Y(\phi)$ only decreases the absolute value of $S_{\rm REE}$, which is consistent with the expected monotonicity property.

\subsection{Numerical Results  for the Polar Caps}\label{subsec-numeric}

We now turn to entangling regions on a spherical cut-off boundary which are circles smaller than the great circle, enclosing a polar cap region of angle $\theta_0 < \pi/2$.  In this case,  the Euler-Lagrange equations obtained by varying the area functional leads to a second-order differential equation which does not admit a simple closed-form solution.  We therefore solve the Euler-Lagrange equations numerically to determine the minimal surface and having obtained this solution,  determine the area by inserting the solution in the area functional.

\begin{figure}
   \centering
        \includegraphics[width=0.7\textwidth]{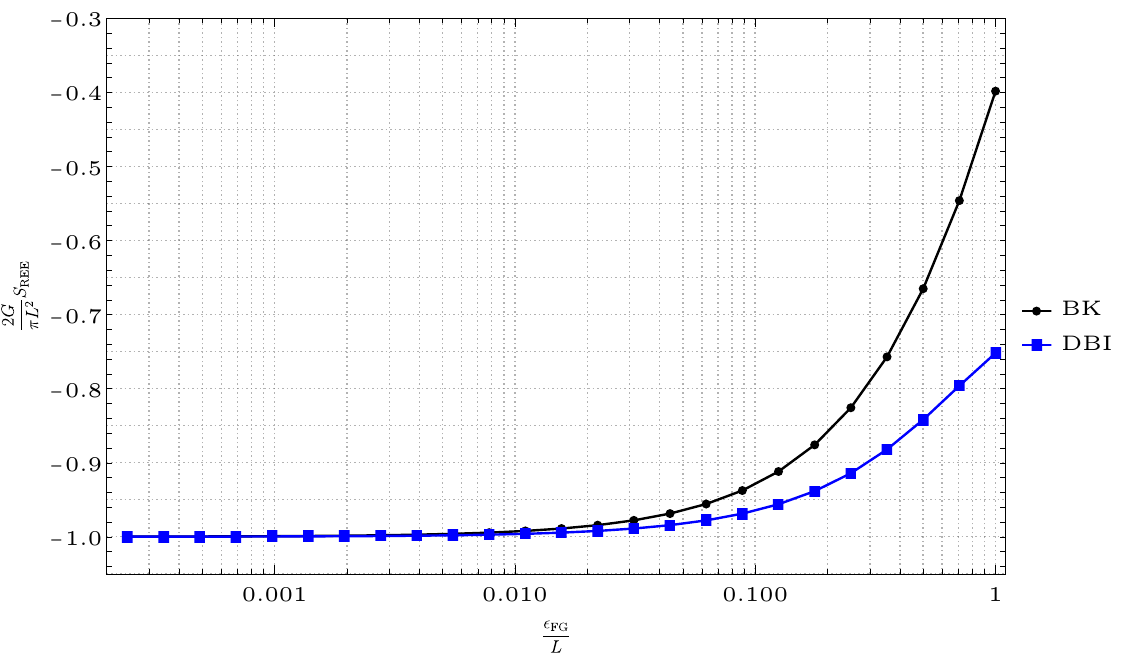}
        \caption{Variation of the renormalized entanglement entropy with the radial FG cut-off $\epsilon_{\mathrm{FG}}/L$ for a fixed angle $\theta_0 = \pi/4$ for the two different counter-terms.}
        \label{fig:plot-a}
\centering
   \includegraphics[width=0.7\textwidth]{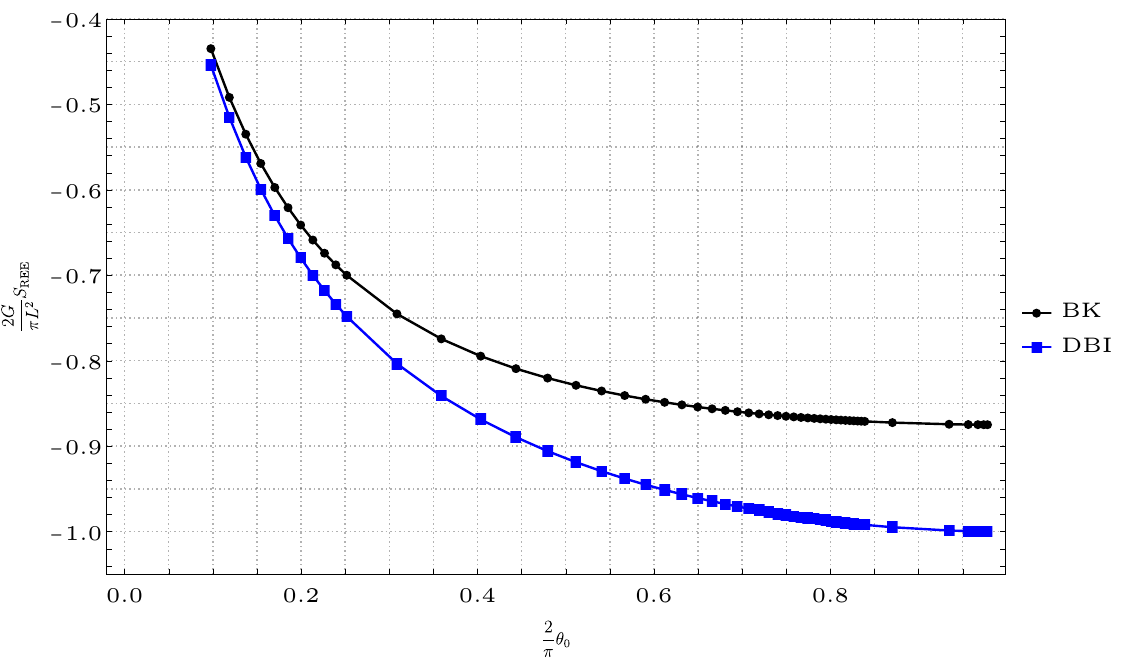}
        \caption{Variation of the renormalized entanglement entropy  with the polar cap angle $\theta_0$ at a fixed radial FG cut-off $\epsilon_{\rm FG}/L = 0.25$. }
        \label{fig:plot-b}
\end{figure}

In the general scenario,  the discussion of holographic renormalization \cite{Skenderis:2002wp} necessitates that we work with the Fefferman-Graham (FG) gauge
\cite{Fefferman:1985con} in which the metric has the form
\baa
\dd{s}^2 = \frac{L^2}{z^2} \bqty{ \dd{z}^2 + g_{\mu \nu} (z, x) \dd{x}^\mu \dd{x}^\nu}, \label{fgmetric}
\ea
where $x^\mu$ denotes the boundary coordinates.
It is meaningful to talk about holographic renormalization in this gauge,  where the finite cut-off corresponds to a cut-off in the $z$-coordinate.  Note that all this subtlety is immaterial for the discussion so far in the paper, because all the radial coordinates used so far are monotonic functions of the $z$ coordinate.  Therefore, all the results so far, including the shape-independence of the $F$-function and the equatorial entanglement entropy would be the same no matter what coordinate we use.  This statement is true for this section as well.  However,  since we want to see how the renormalized entanglement entropy runs under holographic RG flow,  it is more natural to study the entropy as a function of $\epsilon_{\rm FG}$ where the cut-off is located at $z = \epsilon_{\rm FG}$.

The FG $z$ coordinate is related to the radial $r$ coordinate in the geometry \eqref{adsmet} simply as
\baa
z = 2 \qty(\sqrt{r^2 + L^2} - r).  \label{zrrelation}
\ea

In Figure \ref{fig:plot-a},  we study the variation of  $S_{\mathrm{REE}}$ with the radial cut-off at a fixed polar cap angle $\theta_0  = \pi /4$.  The plot is consistent with the well-known result \cite{Casini:2011kv} that in the asymptotic limit the renormalized entanglement entropy $S_{\mathrm{REE}}$ saturates to the value $- \pi L^2 / 2 G$, independent of the angle $\theta_0$. This is the same value obtained for the equatorial RT surface \eqref{sreedisc}.  It is worth noting that  the behavior of the absolute value of the renormalized entropy,  which can be taken as a measure of the number of degrees of freedom,  is consistent with the expected monotonicity properties --- it decreases monotonically under an increase of the cut-off $\epsilon_{\rm FG}$.  It is generally true that at any scale
\baa
| S_{\mathrm{REE, DBI}} | \geq | S_{\mathrm{REE, BK}} |, \label{dbibkrel}
\ea
which implies that the DBI counter-term corresponds to integrating out fewer degrees of freedom than the BK counter-term,  which was also our observation for the $F$-function in \S\ref{sec-s3pf}.

In Figure \ref{fig:plot-b}, we show how the renormalized entanglement entropy depends on the polar cap angle $\theta_0$.  We find that the DBI counter-term,  near the equator saturates to the asymptotic value \eqref{sreedisc} which further confirms our analytic results.  {We carried out the numerical analysis in Mathematica with \texttt{WorkingPrecision} at least as high as 25.  As such,  any  error in the value of $S_{\rm REE}$ is too small to be depicted on the plots. }

\section{Discussion}\label{sec-disc}

In this work,  we have presented a different take on the utility of the DBI counter-term.  We first showed that compared to any other counter-term, the DBI counter-term gives a constant maximum value of the $F$-function for the boundary field theory defined on $S^{3}$.  This, in particular, implies that the $F$-function derived using the DBI counter-term is independent of the radial cut-off.  This led us to speculate that, $F_{\mathrm{DBI}}$, the $F$-function with DBI term, may be a topological invariant.  We established that by considering the radial cut-off to be dependent on the polar angle $\theta$ of $S^{3}$ and demonstrated that the integrand can be written as a total derivative with respect to $\theta$ and an additive piece,  with the total derivative term yielding a vanishing contribution to the final answer.

We then embarked on the computation of the renormalized entanglement entropy using the replica method in gravity \cite{Lewkowycz:2013nqa} refined in \cite{Taylor:2016aoi}.  We established that for the global AdS case,  the $S_{\rm REE}$ for an equatorial surface is independent of the cut-off only if one uses the DBI counter-term.  We then focused on the equatorial entangling surface with the bulk radial cut-off having an arbitrary dependence on the coordinate parametrizing the surface and showed that the $S_{\rm REE}$ is still independent of the shape of the cut-off surface with DBI counter-term for RT surfaces normal to the cut-off.  This result seems to point towards a hidden integrable structure in the deformation of the boundary curve.  We will comment on it momentarily.

The situation is not as easy if one considers asymmetric entangling surfaces,  i.e., if the (circular) surface lies not on the equator but in one of the hemispheres.  This situation is not amenable to analytic methods and we took recourse to performing numerical computations.  We showed that the DBI counter-term is always consistent with integrating out fewer degrees of freedom compared to the BK counter-term.

{Let us make a few more comments about the field theory interpretation of our results. We found that the candidate $F$-function defined holographically for our counter-term does not run with the RG flow,  for azimuthally symmetric cut-off surfaces with an otherwise arbitrary shape.  One possible explanation is that this particular $F$-function corresponds to an exactly marginal deformation of the three-dimensional CFT. Such exactly marginal deformations have been known in theories with a sufficient amount of supersymmetry \cite{Bianchi:2010cx,  Ashmore:2018npi}.  A particularly simple example of an exactly marginal deformation would be that of a free compact scalar in two dimensions, for which $c=1$ for any value of the radius. For the half-sphere, $S_{\rm REE}$ also exhibits a similar behavior. However, $S_{\rm REE}$ for the polar cap, in contrast, runs in a strictly monotonic manner. This suggests the possibility that our candidate $F$-function and $S_{\rm REE}$ for the half-sphere show such a special behavior on account of the underlying symmetry of the geometry and are insensitive to the degrees of freedom being integrated out. It would be interesting to explore these field-theoretic aspects in more detail. }

We will now briefly comment on the possible integrable structure hidden in this analysis.  The entangling surface in the 3D CFT on the boundary of the global AdS$_{4}$,  on the constant time subspace,   is just a closed curve.  The corresponding RT surface in the bulk has the topology of a disk.  For the equatorial entangling surface, the RT surface has the geometry of Euclidean AdS$_{2}$ which is an equal time slice of the AdS$_{3}$.  Usually when the RT surface ends on the boundary entangling surface one chooses the Dirichlet boundary condition; however, in the case of AdS$_{2}$ geometry of the RT surface for the equatorial entangling surface one can, in principle, implement a variety of boundary conditions.  These boundary conditions are, in some sense, pullback of the boundary conditions used in the AdS$_{3}$ case \cite{Perez:2016vqo}.  {It would also be interesting to examine situations with additional matter fields and/or black holes in the bulk --- the results in \cite{Jatkar:2022zdz} are encouraging in this regard.  While a general higher-dimensional analog of the special counter-term considered here is not known, it would be interesting to pursue a similar higher-dimensional generalization.}

The Dirichlet boundary condition corresponds to the celebrated Brown-Henneaux boundary conditions for AdS$_{3}$ \cite{Brown:1986nw}.  Besides this boundary condition, there are an infinite number of other boundary conditions which are governed by two copies of the Korteweg–de Vries (KdV) hierarchy \cite{Perez:2016vqo}.
The AdS$_{2}$ geometry also permits boundary conditions
governed by a single copy of the KdV hierarchy \cite{Cardenas:2024hah}.  It would be interesting to determine the interplay between the choice of boundary conditions and boundary counter-terms and their impact on the computation of various entanglement measures.

More interestingly,  the dynamics of a curve on a plane \cite{Goldstein:1991zz} as well as on two-dimensional sphere \cite{Doliwa:1994bk} is governed by the modified KdV (mKdV) hierarchy.  Note that the entangling surface on the boundary is a closed curve on the two-dimensional boundary sphere.  The infinite number of conserved quantities associated with the mKdV hierarchy could possibly be put to good use in deforming the circular entangling surface into various other shapes consistent with the mKdV conservation laws.  We are pursuing both these aspects of integrability but they are beyond the scope of this work.

\acknowledgments

DPJ and UM acknowledge the hospitality of the Abdus Salam International Centre for Theoretical Physics during the progress of this work.  DPJ would like to acknowledge support from the ICTP through the Associates Programme (2022-2027).  UM's research is supported by the European Research Council under the European Union’s Seventh Framework Programme (FP7/2007-2013), ERC Grant agreement ADG 834878.

\appendix

\section{\boldmath Components of the Ricci Tensor for the Equatorial Cut-Out}\label{app-a}

Here we will spell out the components of the Ricci tensor for the metric \eqref{dfinduced} on the equatorial cut-out; see \S\ref{subsec-rtcut}. In the case of a constant radial cut-off, the $tt$ component of the Ricci tensor is zero and therefore the curvature along this direction does not contribute to the final answer.  However,  when we have an arbitrary cut-out surface,  this component has a non-zero value,  as can be seen in the following expression.  Nevertheless,  as the cut-off surface is static in a static spacetime,  the $R_{t\theta} = 0  = R_{t \phi}$.  Suppressing the explicit dependence on the angle $\phi$,  the non-vanishing independent components are given by:

\baa
R_{tt} &= - \frac{(1+ X^2)}{L^2 \left(X^4+X^2+\left(X'\right)^2+Y^2\right)^2 } \Bigg[ 2 \left(\left(X'\right)^2+Y^2\right)^2+X^5 \left(X''+Z\right)\\
 &\quad +X^3 \left(X''+Z\right) +X^2 \left(\left(X'\right)^2+Y^2\right) +X \left(Z \left(X'\right)^2+Y \left(Y X''-2 X' Y'\right)\right) \Bigg], \label{rtt}
\ea

\baa
R_{\theta \phi} &=  - \frac{1}{X \left(X^2+1\right) \left(X^4+X^2+\left(X'\right)^2+Y^2\right)^2}  \Bigg[ X^{10} Y'+2 X^8 Y'-2 X^9 Y X' \\
&\quad  -X^5 Y X' \left(2 \left(X'\right)^2+2 Y^2+3\right)+2 Y X' \left(Z \left(X'\right)^2+Y \left(Y X''-2 X' Y'\right)\right)\\
&\quad +X^2 Y X' \left(\left(3 Y^2+1\right) X''-6 Y X' Y'+3 Z \left(X'\right)^2+Z\right)\\
&\quad +X Y X' \left(-Z X''+Y^2 \left(4 \left(X'\right)^2-2\right)+2 \left(X'\right)^4-2 \left(X'\right)^2+2 Y^4+\left(Y'\right)^2\right)\\
&\quad +X^6 \left(\left(\left(X'\right)^2+1\right) Y'+Y X' \left(X''+Z\right)+Y^2 Y'\right)\\
&\quad +X^4 \left(\left(X'\right)^2 Y'+2 Y X' \left(X''+Z\right)+Y^2 Y'\right)\\
&\quad -X^3 Y X' \left(Z X''+3 \left(X'\right)^2+3 Y^2-\left(Y'\right)^2+1\right)  -4 X^7 Y X'\Bigg], \label{rthetaphi}
\ea

\baa
R_{\theta \theta} &=  \frac{1}{2 X \left(X^2+1\right) \left(X^4+X^2+\left(X'\right)^2+Y^2\right)^2} \Bigg[ 2 X^{11}+2 X^9 \left(2 Y^2+3\right)\\
&\quad -2 X^{10} \left(X''+2 Z\right)-2 X^8 \left(3 X''+5 Z\right) -4 Y^2 \left(Z \left(X'\right)^2+Y \left(Y X''-2 X' Y'\right)\right) \\
&\quad -2 X^2 \left(Y^2 \left(3 \left(Y^2+1\right) X''+Z\right)+\left(3 Y^2+2\right) Z \left(X'\right)^2-2 \left(3 Y^2+2\right) Y X' Y'\right)\\
&\quad +X Y^2 \left(2 Z X''+\left(4-8 Y^2\right) \left(X'\right)^2-4 \left(X'\right)^4-4 Y^4+4 Y^2-2 \left(Y'\right)^2\right) \\
&\quad +2 X^7 \left(Z X''+2 \left(X'\right)^2+6 Y^2- \left(Y'\right)^2+3\right) \\
&\quad -2 X^6 \left(\left(4 Y^2+3\right) X''+2 Z \left(2 \left(X'\right)^2+Y^2+2\right)-6 Y X' Y'\right)\\
&\quad +X^5 \left(4 Z X''-4 \left(Y^2-2\right) \left(X'\right)^2-4 \left(X'\right)^4+14 Y^2-4 \left(Y'\right)^2+2\right) \\
&\quad -2 X^4 \left(\left(7 Y^2+1\right) X''+Z \left(6 \left(X'\right)^2+3 Y^2+1\right)-10 Y X' Y'\right)\\
&\quad +2 X^3 \left(Z X''-2 \left(X'\right)^4+2 \left(X'\right)^2-Y^2 \left(-Z X''+\left(X'\right)^2+\left(Y'\right)^2-3\right) \right) \\
&\quad + 2 X^3\left(Y^4-\left(Y'\right)^2\right) \Bigg], \label{rthetatheta}
\ea

\baa
R_{\phi \phi} &=  \frac{1}{X \left(X^2+1\right) \left(X^4+X^2+\left(X'\right)^2+Y^2\right)^2}  \Bigg[ X^{11}-X^{10} \left(2 X''+Z\right)\\
&\quad -X^8 \left(5 X''+3 Z\right)-2 \left(X'\right)^2 \left(Z \left(X'\right)^2+Y \left(Y X''-2 X' Y'\right)\right) \\
&\quad +X^2 \left(2 Y X' \left(3 \left(X'\right)^2+2\right) Y'-3 Z \left(\left(X'\right)^4+\left(X'\right)^2\right)\right) +X^9 \left(2 \left(X'\right)^2+3\right)\\
&\quad -X \left(X'\right)^2 \left(-Z X''+Y^2 \left(4 \left(X'\right)^2-2\right)+2 \left(X'\right)^4-2 \left(X'\right)^2+2 Y^4+\left(Y'\right)^2\right)\\
&\quad +X^7 \left(Z X''+6 \left(X'\right)^2+2 Y^2-\left(Y'\right)^2+3\right)\\
&\quad +X^6 \left(6 Y X' Y'-Z \left(4 \left(X'\right)^2+3\right)-2 X'' \left(\left(X'\right)^2+2 Y^2+2\right)\right)\\
&\quad +X^5 \left(2 Z X''-2 Y^2 \left(\left(X'\right)^2+Y^2-2\right)+7 \left(X'\right)^2-2 \left(Y'\right)^2+1\right)\\
&\quad -X^4 \left(-10 Y X' Y'+7 Z \left(X'\right)^2+X'' \left(3 \left(X'\right)^2+6 Y^2+1\right)+Z\right)\\
&\quad +X^3 \left(Z X''-Y^2 \left(\left(X'\right)^2-2\right)+\left(X'\right)^2 \left(Z X''+\left(X'\right)^2-\left(Y'\right)^2+3\right)  \right) \\ &\quad -X^3 \left( 2 Y^4 + \left(Y'\right)^2\right) - X^2 X'' \left(\left(3 Y^2+1\right) \left(X'\right)^2+2 Y^2\right) \Bigg].\label{rphiphi}
\ea

\bibliographystyle{JHEP}
\bibliography{DBI-RT}

\end{document}